\documentclass[%
reprint,
amsmath,amssymb,
pra,
longbibliography,
]{revtex4-2}
\usepackage[utf8]{inputenc}
\usepackage{graphicx}% Include figure 
\usepackage{dcolumn}% Align table columns on decimal point
\usepackage{bm}% bold math
\usepackage[english]{babel}
\usepackage{enumitem}
\usepackage{xcolor}
\usepackage{hyperref}
\hypersetup{
	colorlinks=true,
	linkcolor=blue,    % color of internal links
	citecolor=blue,   % color of links to bibliography
	filecolor=magenta, % color of file links
	urlcolor=blue      % color of external links
}

\begin{document}

\preprint{APS/123-QED}

%\title{Spin-Resolved Electron Diffraction from Nanogratings: Self-Field Effects and Magnetic Control}
\title{Magnetically assisted spin-resolved electron diffraction: Coherent control of spin population and spatial filtering}

\author{Sushanta Barman}
\email{sushanta.ice@gmail.com}
\author{Kuldeep Godara}
\email{kuldeep22@iitk.ac.in}
\author{Sudeep Bhattacharjee}%
\email{sudeepb@iitk.ac.in}
\affiliation{Department of Physics, Indian Institute of Technology-Kanpur, Kanpur 208016, India}%

%======================================= ABSTRACT ========================
\begin{abstract}
Electron diffraction from nanogratings provides a platform for free-electron interferometry, yet controlled manipulation of the electron spin within such diffraction geometries remains largely unexplored. In particular, the role of the self-generated magnetic field due to the motion of an electron and the feasibility of coherent spin control without disrupting diffraction coherence have not been quantitatively investigated. In this article, a self-consistent Maxwell-Pauli framework is developed to study spin-resolved electron diffraction from nanogratings in the presence of magnetic fields. The model incorporates geometric confinement, image-charge interactions, self-generated magnetostatic fields, and externally applied magnetic fields. Numerical simulations show that the intrinsic magnetic self-field produced by the electron probability current is several orders of magnitude too weak to induce measurable spin mixing, demonstrating that nanogratings act as spin-conserving beam splitters in the absence of external magnetic fields. When a uniform magnetic field is applied upstream of the grating, coherent Larmor precession enables controlled spin rotation without having to modify the diffraction geometry. The $\pi$-rotation field scales inversely with the interaction length and the electron de Broglie wavelength $\lambda_{dB}$. A downstream nonuniform magnetic field applied after the nanograting, imparts a spatially varying Zeeman phase, thereby generating opposite transverse momentum shifts for the two spin components. In addition, the spin-dependent transverse dynamics is analyzed using Husimi Q-function phase space maps, which provide a direct phase space visualization of population redistribution and spin-dependent momentum shifts. The proposed method produces tunable spatial separation of spin-resolved electrons. The result demonstrates an all-magnetic route to coherent spin rotation and analysis in free-electron diffraction and provide a quantitative basis for spin-resolved free-electron generation and interferometry.
\end{abstract}

\maketitle

%\tableofcontents

%===========================================================================
%                          INTRODUCTION
%===========================================================================
\section{\label{introduction}Introduction}

Electron interferometry offers a direct probe of quantum coherence and phase evolution in free space, where precise control over both spatial and spinor degrees of freedom is essential for applications ranging from electron microscopy to quantum metrology and sensing \cite{Ruimy2025, Priebe2017, 8f7h-6nfc}. Since the pioneering biprism experiments of Möllenstedt and Düker \cite{Moellenstedt1956}, nanofabricated beam splitters have enabled a wide range of phase-sensitive electron-optical techniques, including holography \cite{Hasselbach_2010, PhysRevLett.74.399}, electromagnetic field mapping \cite{RevModPhys.59.639, 1nfc-stxp}, and aberration-corrected imaging \cite{10.1093/jmicro/dfaa033}. In particular, free-standing nanogratings have emerged as reliable amplitude-dividing beam splitters for electron matter waves, which enable Lau, Talbot, and Mach–Zehnder interferometers \cite{McMorran_2009,PhysRevA.74.061602,10.1063/1.2357000,PhysRevLett.127.110401, PhysRevResearch.3.043009, PhysRevLett.126.146803}.

Despite these advances, electron diffraction using nanogratings has been mostly explored in the context of spatial coherence \cite{10.1063/1.2357000}. The controlled manipulation of the electron spin in such diffraction geometries remains largely unexplored. A diffraction-based platform that preserves spatial coherence and provides coherent spin control would allow the study of spin-dependent phase accumulation, spin-orbit coupling, and exchange interactions at the nanoscale. Such capabilities are essential for spin-resolved interferometry \cite{PhysRevLett.118.070403, PhysRevB.95.085411}, quantum state tomography \cite{Cramer2010, PhysRevLett.109.120403}, and nanoscale magnetic-field sensing \cite{Marchiori2022, Maze2008}, with direct relevance to spintronics and spin-contrast electron microscopy.

A central challenge in realizing spin-resolved electron diffraction is achieving magnetic control of the spin degree of freedom. Previous studies of spin-dependent electron scattering have focused primarily on surface-sensitive or solid-state systems. Spin-polarized low-energy electron diffraction probes magnetic order at crystalline surfaces through spin-orbit and exchange interactions \cite{R_Feder_1981}, while spin-spiral interfaces induce spin-dependent diffraction and interfacial spin currents \cite{10.1063/1.2837479}. In parallel, strong-field and relativistic studies have examined spin forces and radiation-reaction effects for free electrons in intense laser fields \cite{PhysRevA.95.042102, Blackburn2020, Mishra2022}. However, none of these approaches address coherent spin evolution in a free-space electron diffraction geometry defined by nanostructured beam splitters, and furthermore, separation of electron spin states from a mixed state by employing magnetic tuning.

Moreover, the influence of the electron probability current and the associated magnetostatic self-field on spinor phase evolution during diffraction has not been quantified. A comprehensive theoretical framework is therefore required that combines realistic grating potentials, self-consistent magnetostatics, and externally applied magnetic fields within a unified description of spin-resolved electron diffraction.

In this work, spin-resolved electron diffraction from a free-standing nanograting is investigated using a self-consistent Maxwell-Pauli framework. The system consists of low-energy (20~eV) electrons diffracted by a multi-slit transmission grating, where the electron wave function is described as a two-component wave packet propagating in free space. The electron beam can be extracted from a microwave plasma-based source confined by a magnetic multicusp field \cite{10.1063/1.2764445, 10.1063/1.2943341, 10.1063/1.3117527, 10.1063/1.3369287}. Such sources provide efficient electron production with high electron densities, typically \( n_e \sim 10^{11}\,\mathrm{cm^{-3}} \) \cite{10.1063/1.2943341}. Following extraction, the electron beam can be collimated using appropriate charged-particle optics to meet the experimental requirements. The grating is modeled using a geometric confinement potential together with an image-charge interaction that accounts for electron-metal surface interactions. Spin dynamics are governed by the Pauli Hamiltonian, including contributions from externally applied magnetic fields and the magnetostatic self-field generated by the electron probability current. The coupled spinor evolution and magnetostatics are solved numerically using a split-step Fourier method with self-consistent updating of the magnetic vector potential. This approach allows the intrinsic self-field effects to be quantitatively separated from those due to external magnetic fields. It is shown that the self-generated magnetic field remains negligibly small and does not induce measurable spin mixing during diffraction. Controlled spin manipulation is instead achieved by applying a uniform magnetic field $\mathbf{B}_1$ in front of the grating to induce coherent Larmor rotation, followed by a nonuniform magnetic field $\mathbf{B}_2$ behind the grating that imparts a spatially varying Zeeman phase. In addition, the spin-dependent transverse dynamics are analyzed using spin-resolved Husimi $Q$-function phase-space maps, which provide a direct phase-space visualization of population redistribution and spin-dependent momentum shifts. This mechanism enables tunable spatial separation of the spin-resolved electrons and their diffraction patterns. These results provide a quantitative framework for spin-dependent electron diffraction and coherent spin control using nanogratings, and support spin-resolved free-electron interferometry and the development of related beam-optical devices.

The article is structured as follows: Section \ref{methodology} explains the theoretical methods. Section \ref{result_and_discussions} presents and discusses the simulation results. Finally, Section \ref{conclusions} draws the conclusions.

\begin{figure}
	\centering	\includegraphics[scale=0.95]{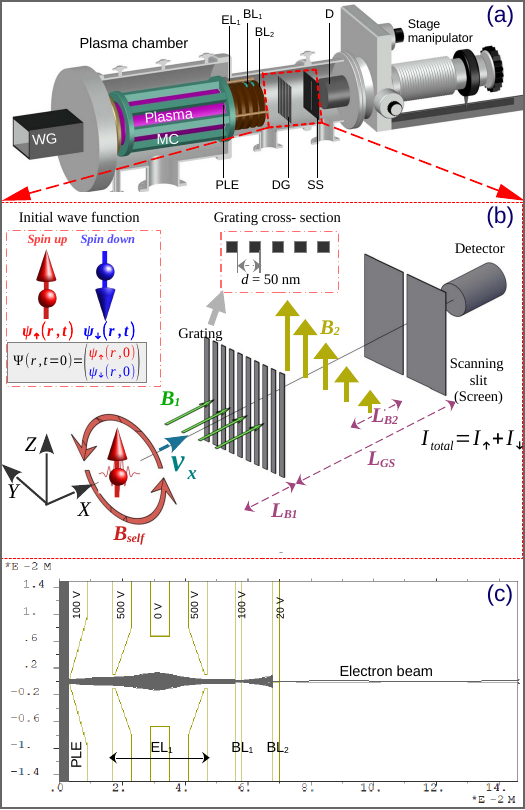}
	\caption{(a) Schematic diagram of the experimental system. WG: waveguide; MC: multicusp; PLE: plasma electrode; EL$_1$: Einzel lens system; BL$_1$ and BL$_2$: beam collimating apertures; DG: diffraction grating; SS: scanning slit; and D: detector. The electron beam is extracted from the plasma. (b) Schematic of the simulation setup. Spin-polarized electrons with kinetic energy $E=20$ eV, initially aligned along the $z$ axis, propagate along $x$ and encounter a nanograting of period $d=50$ nm and thickness $h=25$ nm placed in the $yz$ plane. Upstream of the grating, electrons pass through a uniform magnetic field $\mathbf{B}_{1} = B_{1}\hat{\mathbf{x}}$ over a length $L_{B1}$, followed downstream by a nonuniform field $\mathbf{B}_{2} = zG_{2}\hat{\mathbf{y}} + yG_{2}\hat{\mathbf{z}}$ extending over $L_{B2}$. The far-field diffraction pattern is recorded on a screen at a distance $L_{GS}$ using a scanning slit and detector. (c) Electron trajectory simulation results obtained using AXCEL-INP with plasma parameters as input. The given electrode configuration and applied voltages produce a well-collimated electron beam.}   \label{schematic_diagram_of_simulation_setup}
\end{figure}
%===========================================================================
%                          METHODS
%===========================================================================
\section{\label{methodology} Methodology}
\subsection{Physical model and simulation setup}
The schematic of the experimental setup is shown in Fig. \ref{schematic_diagram_of_simulation_setup}(a), which consists of a microwave (2.45 GHz) driven multicusp plasma source, a plasma electrode (PLE), a beam extraction Einzel lens (EL$_1$), two beam collimating lenses (BL$_1$ and BL$_2$), the diffraction grating (DG), the scanning slit (SS), and the detector (D). The expanded view of the simulation region is shown in Fig. \ref{schematic_diagram_of_simulation_setup}(b). Electrons propagate along the $x$ axis and encounter a transmission grating of period $d = 50~\mathrm{nm}$, open fraction $f = 50\%$, and thickness $h=25~\mathrm{nm}$, positioned in the $yz$ plane, as shown in Fig. \ref{schematic_diagram_of_simulation_setup}(b). To investigate the effect of external magnetic fields on the diffraction pattern, electrons first traverse a uniform magnetic field region, $\mathbf{B}_{1} = B_{1}\hat{\mathbf{x}}$ (for $0\leq x \leq L_{B1}$, where the origin 0 is taken as the start of the magnetic field region) of length $L_{B1}$ upstream of the grating, inducing spin rotation due to Larmor precession. After the grating, they pass through a nonuniform magnetic field $\mathbf{B}_{2} = zG_{2}\hat{\mathbf{y}} + yG_{2}\hat{\mathbf{z}}$ (for $L_{B1} + (L_{GS}-L_{B2})/2 \leq x \leq L_{B1} + (L_{GS}+L_{B2})/2$) of length $L_{B2}$, which spatially separates the diffraction fringes corresponding to the two spin components. A detection screen is placed at $L_{GS} = 50$ cm downstream of the grating to record the far-field diffraction pattern.

The quantum state of a spin-polarized electron is represented by a two-component spinor \cite{PhysRevResearch.5.023164, PhysRevA.96.052132},
\begin{align}
\Psi(\mathbf{r},t) =
\begin{pmatrix}
\psi_{\uparrow}(\mathbf{r},t) \\
\psi_{\downarrow}(\mathbf{r},t)
\end{pmatrix},
\end{align}
where \(\psi_{\uparrow}(\mathbf{r},t)\) and \(\psi_{\downarrow}(\mathbf{r},t)\) are the spatial components associated with the eigenstates of the Pauli operator
\(\sigma_z = \begin{pmatrix} 1 & 0 \\ 0 & -1 \end{pmatrix}\),
corresponding to spin aligned along \(+\hat{z}\) (spin-up) and \(-\hat{z}\) (spin-down), respectively. The populations of the two spin components are given by \(P_{\uparrow,\downarrow}(t)=\int d^2r\,|\psi_{\uparrow,\downarrow}(\mathbf r,t)|^2\),
which represent the probabilities of finding the electron in the corresponding spin state. The total probability density is given by \(\rho(\mathbf{r},t)=|\Psi(\mathbf{r},t)|^{2} =|\psi_{\uparrow}(\mathbf{r},t)|^{2} +|\psi_{\downarrow}(\mathbf{r},t)|^{2}\). 

The electrons are extracted from a multicusp plasma source using a combination of a plasma electrode PLE (circular aperture size 260 $\mu$m) kept at 100 V and an extraction electrode EL$_1$ at 500 V. The typical plasma potential lies in the range 30–40 V. The extracted beam is collimated using a set of two more electrodes BL$_1$ (circular aperture size 100 $\mu$m) and BL$_2$ (circular aperture size 50 $\mu$m), with decelerating potentials of 100 V and 20 V, respectively, applied to them, and the beam kinetic energy is reduced to about 20 eV. Fig. \ref{schematic_diagram_of_simulation_setup}(c) shows an AXCEL-INP simulation of the extracted beam trajectory from the plasma source with the potentials applied to the different electrodes. The geometrical details of PLE, EL$_1$, and BL$_1$ are provided in Refs.~\cite{10.1063/1.5094511,Barman_2022,Barman_2024}. In the present study, EL$_2$ of the previous configuration is replaced by BL$_2$ (thickness 2 mm), positioned 10 mm after BL$_1$. The plasma parameters, such as the plasma potential (40 V), the electron temperature (10 eV), ion temperature (0.2 eV), the electron mass in amu (0.0005486), and the electron current density (40.1 A/m\textsuperscript{2}) at the PLE aperture, are provided as input to the AXCEL-INP code \cite{spadtke1999axcel}. For a 20 eV beam, the fringe width $\gamma$ ( $= L_{GS}\lambda_{dB}/d) \approx 2.75$ mm. This can be well resolved by the $X$–$Y$ stage controller (V. G. Scienta), which has a resolution of 0.5 $\mu$m. The detector D is a channeltron (Surface concept GmbH, CEM4230 R004) which is moved along with the scanning slit SS (Edmund Optics, aperture size : 5 micron $\times$ 3 mm) in the $y$-direction parallel to the grating DG. The particle counts are recorded as a function of $y$ to measure the diffraction pattern.  In the present article, the diffraction of the spin-polarized electron beam depicted in the region as shown in Fig. \ref{schematic_diagram_of_simulation_setup}(b) is investigated through numerical simulations.

For electrons with a kinetic energy of $20$ eV, the velocity $v/c \approx 0.0088 \ll 1$, is well within the nonrelativistic regime; hence we employ the Schrödinger-Pauli formalism. In this limit, the Dirac equation reduces to the Pauli Hamiltonian, which captures orbital motion via minimal coupling and spin dynamics through the Zeeman interaction. The time-dependent Schrödinger-Pauli equation in two dimensions $(x,y)$ for the two-component spinor wave function $\Psi(x,y,t)$ is given by \cite{PhysRev.78.29,PhysRevA.96.052132, RevModPhys.65.733},
\begin{align}
i\hbar\,\frac{\partial\Psi(x,y,t)}{\partial t}
&=
\Bigg[
\frac{1}{2m_e}\big\{-i\hbar\nabla + e\,\mathbf A(x,y,t)\big\}^2 \nonumber \\
&\quad + V_g(x,y) + V_{\mathrm{image}}(x,y) \nonumber \\
&\quad - \frac{g\mu_B}{2}\,\boldsymbol{\sigma}\!\cdot\!\mathbf B(x,y,t)
\Bigg]\Psi(x,y,t),
\label{eq:PSH}
\end{align}
where $\hbar$ is Planck’s constant, $m_e$ the electron mass, $e$ the elementary charge, $\mu_B = e\hbar/(2m_e)$ the Bohr magneton, $g = -2.00231930436256(35)$ is the Lande $g$-factor \cite{RevModPhys.93.025010}, and $\boldsymbol{\sigma}$ denotes the Pauli matrices. The scalar potential energies $V_g(x,y)$ and $V_{\mathrm{image}}(x,y)$ represent the geometric confinement of the wave packet due to the multi-slit grating and the image-charge interaction of the electron with the conducting bars of the grating, respectively. The magnetic field $\mathbf{B}(x,y,t)$ and the vector potential $\mathbf{A}(x,y,t)$ include contributions from the externally applied magnetic fields ($\mathbf{B}_{1}$ and $\mathbf{B}_{2}$) as well as the self-field $\mathbf{B}^{\text{self}}$ generated by the electron probability current. The explicit expansions of the Hamiltonian before and after the grating, denoted by $H_1$ and $H_2$, are presented in Appendices~\ref{app:PauliExpansion_B1} and~\ref{app:PauliExpansion_B2}, respectively.

\begin{figure}
	\centering
	\includegraphics[scale=0.95]{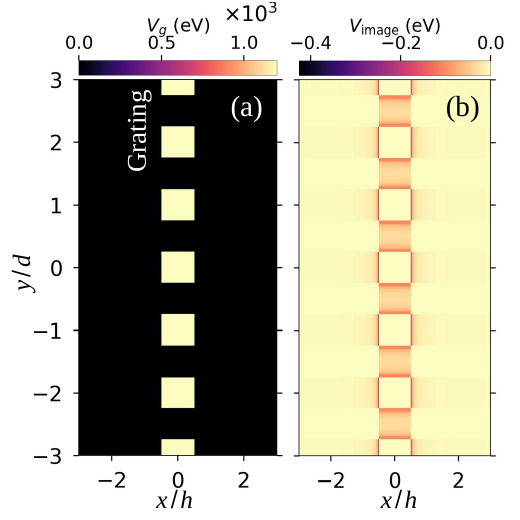}
	\caption{Spatial maps of (a) the geometric potential energy $V_g(x,y)$ and (b) the image-charge potential energy $V_{\mathrm{image}}(x,y)$ for an electron in a multi-slit grating.}
    \label{V_g_and_V_image}
\end{figure}

\subsection{Geometric potential energy $V_{g}(x,y)$ and image-charge potential energy $V_{\mathrm{image}}(x,y)$}

The grating is modeled using two potential energies, the geometric potential energy \(V_{g}(x,y)\) and the image-charge potential energy \(V_{\mathrm{image}}(x,y)\), both in units of eV. The geometric potential energy \(V_{g}(x,y)\) describes the confinement due to the grating structure and is implemented as a step potential \cite{Barman_2023, 10.1063/5.0098030},
\begin{equation}
V_g(x,y) =
\begin{cases}
0, & |y - y_c| \leq w/2 \ \ \text{(inside slit openings)}, \\[6pt]
V_0, & \text{otherwise},
\end{cases}
\label{eq:Vgeo}
\end{equation}
where $w = d/2$ is the slit width, $y_c$ is the center of the slit, and $V_0 = 1200\,\mathrm{eV}$ is chosen to be much larger than the electron kinetic energy ($E_0 = 20$ eV), thereby suppressing quantum tunneling through the opaque grating bars \cite{Barman_2023}. The distribution of \(V_{g}(x,y)\) is shown in Fig.~\ref{V_g_and_V_image}(a).

\begin{figure}[b]
\centering	\includegraphics[scale=0.95]{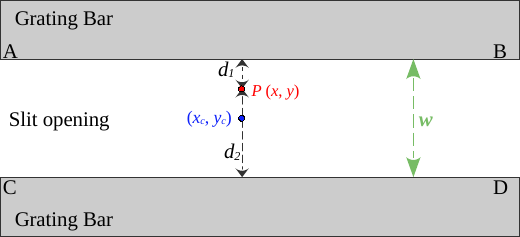}
	\caption{Cross-sectional view of the diffraction grating. ABCD represents a slit opening, with AB and CD denoting the planes of two neighboring bars. The distances from a point $P(x,y)$ inside the slit to the AB and CD planes are labeled $d_1$ and $d_2$, respectively.}
    \label{schematic_V_image}
\end{figure}

The image-charge potential energy $V_{\mathrm{image}}(x,y)$ arises from the attractive interaction between the electron and the conducting grating bars, which modifies the electron dynamics. At a point $P(x,y)$ inside a slit opening, as shown in Fig.~\ref{schematic_V_image}, $V_{\mathrm{image}}(x,y)$ is given by \cite{10.1063/1.2357000},
\begin{equation}
V_{\mathrm{image}}(x,y) =
-\frac{\eta\,e^2}{8\pi\epsilon_0}
\left(
\frac{1}{d_1(x,y)} + \frac{1}{d_2(x,y)}
\right),
\label{eq:Vimage}
\end{equation}
where $d_1$ and $d_2$ are the distances from the point $P(x,y)$ to the upper (AB) and lower (CD) edges of the slit, respectively (see Fig.~\ref{schematic_V_image}). The parameter $\eta = 0.35$ is an effective scaling factor that accounts for the reduced strength of the image charge compared to the bare electron charge \cite{10.1063/1.2357000}, and $\epsilon_0$ is the permittivity of free space. This reduction arises from multiple effects. Successive reflections of image charges between parallel walls weaken the effective strength by a logarithmic factor $\ln(2)$, while image formation in dielectric materials such as $\mathrm{Si_3N_4}$ introduces an additional reduction proportional to $\chi_{e}/(\chi_{e}+2)$, where $\chi_{e}$ ($= 8$) is the electric susceptibility \cite{10.1063/1.2357000}. These effects are collectively incorporated into the scaling factor $\eta$, ensuring that the modeled potential realistically represents the electron–surface interaction within the slit region. The spatial distribution of $V_{\mathrm{image}}(x,y)$ is shown in Fig.~\ref{V_g_and_V_image}(b). 
%#############################################################################

\subsection{Self-field magnetostatics}

The charge-current density associated with a two-component Pauli spinor $\Psi(x,y,t)$ can be expressed in terms of its spin-summed components \cite{10.1119/1.19149,Wilkes_2020}:
\begin{align}
J_x =
-e \Bigg[
  \frac{\hbar}{m_e}\sum_{s=\uparrow,\downarrow}\Im\!\big(\psi_s^{*}\,\partial_x\psi_s\big)
  &-\frac{e}{m_e}\,A_x\,\sum_s|\psi_s|^2
  \nonumber \\ &+\frac{\hbar}{2m_e}\,(\nabla\times\mathbf S)_x
\Bigg], \label{eq:Jx}
\end{align}

\begin{align}
J_y =
-e\!\Bigg[
  \frac{\hbar}{m_e}\sum_{s=\uparrow,\downarrow}\Im\!\big(\psi_s^{*}\,\partial_y\psi_s\big)
  &-\frac{e}{m_e}\,A_y\,\sum_s|\psi_s|^2
  \nonumber \\ &+\frac{\hbar}{2m_e}\,(\nabla\times\mathbf S)_y
\Bigg], \label{eq:Jy}
\end{align}
where, the first terms in Eqs.~(\ref{eq:Jx}) and (\ref{eq:Jy}) correspond to the paramagnetic (or convection) current, the second terms to the diamagnetic current associated with the magnetic vector potential, and the third terms to the spin–magnetization current $\nabla\times\mathbf{M}$, where $\mathbf{M} = (e\hbar/2m_{e})\mathbf{S}$ \cite{10.1119/1.19149,Wilkes_2020}. The latter accounts for the contribution of spin to the total charge-current density. The total magnetic vector potential is written as \(\mathbf{A} = \mathbf{A}_{1} + \mathbf{A}_{2} + \mathbf{A}^{\mathrm{self}}\), with \(\mathbf{A}_{1}\) ($=yB_1 \mathbf{\hat{z}}$) and \(\mathbf{A}_{2}\) ($=-(y^2 G_2/2) \mathbf{\hat{x}}$) arising from the externally applied magnetic fields \(\mathbf{B}_{1}\) and \(\mathbf{B}_{2}\), respectively. The term \(\mathbf{A}^{\mathrm{self}}\) denotes the magnetic vector potential generated by the self-field \(\mathbf{B}^{\mathrm{self}}\), which originates from the electron probability current. The quantity $\mathbf{S} = \Psi^{\dagger}\boldsymbol{\sigma}\Psi$ represents the local spin density.

To compute \(\mathbf A^{\mathrm{self}}\), we solve the magnetostatic Poisson equation ($\nabla^2 \mathbf {A}^{\mathrm{self}} = -\mu_0 \mathbf{J}$) in Fourier space within the Coulomb gauge. 
For the nonrelativistic electron velocity (\(v/c \ll 1\)), retardation effects and displacement currents are negligible, and the self-field evolves quasi-statically. Denoting \(\tilde{\mathbf J}(\mathbf k)\) as the Fourier
transform of the charge-current density, the transverse projector
\(P_T(\mathbf k)=\mathbb I-\hat{\mathbf k}\hat{\mathbf k}^{\!\top}\) enforces
\(\nabla\!\cdot\!\mathbf A^{\mathrm{self}}=0\) and yields \cite{ALBERT2022108401}
\begin{equation}
\tilde{\mathbf A}^{\mathrm{self}}(\mathbf k)
=\mu_0\,\frac{P_T(\mathbf k)\,\tilde{\mathbf J}(\mathbf k)}{k^2},
\qquad k^2=k_x^2+k_y^2,
\label{mag_poisson_fourier}
\end{equation}
with the zero mode regularized numerically by setting \(k^2(\mathbf 0)=\infty\). The inverse Fourier transform gives \(A^{\mathrm{self}}_x\) and \(A^{\mathrm{self}}_y\),
and the out-of-plane self-field follows from
\(B^{\mathrm{self}}_z=(\nabla\times\mathbf A^{\mathrm{self}})_z=\partial_xA^{\mathrm{self}}_y-\partial_yA^{\mathrm{self}}_x\)
\cite{ALBERT2022108401}.

%$$$$$$$$$$$$$$$$$$$
\subsection{Time evolution scheme}

Temporal propagation of \(\Psi(x,y,t)\) in Eq.~(\ref{eq:PSH}) is performed using the split-step Fourier method \cite{Barman_2025, PhysRevResearch.6.023165}. 
The Hamiltonian is separated into potential-like contributions and a kinetic part, and the evolution over each time step \(\Delta t\) proceeds via a second-order Strang splitting:
(i) a half-step with the scalar potentials \(V_g(x,y)\) and \(V_{\mathrm{image}}(x,y)\), the Zeeman interaction with the external field \(\mathbf B^{\mathrm{ext}}\) and the self-field \(\mathbf B^{\mathrm{self}}\), and the minimal-coupling terms \(\mathbf A\!\cdot\!\mathbf p\) and \(\mathbf A^2\), where $\mathbf{p}=-i\hbar \mathbf{\nabla}$;
(ii) a full-step kinetic evolution in Fourier space using the exact spectral propagator \(\exp[-i\hbar \Delta t (k_x^2+k_y^2)/(2m_e)]\);
and (iii) a final half-step with the same potential-like terms.

At each discrete time step $t_n=n\Delta t$, where $n$ is an integer ($= 0,1,2,..$), the Maxwell--Pauli coupling is updated self-consistently. Starting from the spinor $\Psi^n$, the charge-current density $\mathbf J[\Psi^n]$ is evaluated in real space using Eqs.~(\ref{eq:Jx}) and (\ref{eq:Jy}). The self-generated vector potential is initialized to zero and is incorporated self-consistently from the subsequent time steps once the charge-current density has been evaluated. The corresponding self-generated vector potential $\mathbf A^{\mathrm{self}}$ is obtained by solving the magnetostatic Poisson equation in Fourier space (Eq.~(\ref{mag_poisson_fourier})), from which the self-field $\mathbf B^{\mathrm{self}}=\nabla\times\mathbf A^{\mathrm{self}}$ is constructed and included in the subsequent split-step propagation together with $\mathbf B^{\mathrm{ext}}$.

Spatial derivatives in the probability current and in the \(\mathbf A\!\cdot\!\mathbf p\) terms are evaluated using centered finite differences, while Fourier methods are used for the kinetic step and the self-field computation.
After each full time step, an absorbing mask \(\Gamma(x,y)\) is applied near the boundaries, as implemented in Ref.~\cite{Barman_2023}, to suppress unphysical reflections.

%#############################################################################
\subsection{Far-field mapping of spin-resolved diffraction patterns}
The far-field distribution of the electron wave packet at the detection screen is obtained under the assumption of paraxial ballistic drift \cite{PhysRevResearch.6.023165}. A transverse momentum component $k_y$ maps to the screen coordinate as $y_{\rm scr}=\frac{\hbar k_y}{m_e} T_{scr}$, where $T_{scr}$ ($=L_{GS}/v$) is the drift time. The intensity at the screen for different spin channels is given by \cite{PhysRevResearch.6.023165},
\begin{equation}
I_s(y_{\rm scr}) = \frac{m_e}{\hbar T_{scr}}\, \big|\tilde\psi_s(k_y)\big|^2,
\qquad s=\uparrow,\downarrow,
\label{eq:I_scr_y}
\end{equation}
where $\tilde\psi_s(k_y)$ is the momentum-space wave function of spin component $s$, obtained from the Fourier transform of the transmitted state $\psi_s(y)$ just after the diffraction,
$\tilde\psi_s(k_y)=\tfrac{1}{\sqrt{2\pi}}\int \psi_s(y)\,e^{-ik_y y}\,dy$.

For spin readout in the \(\sigma_y\) basis, the wave function is projected onto the eigenstates
\(|\!\pm y\rangle=(|\uparrow z\rangle \pm i|\downarrow z\rangle)/\sqrt{2}\) \cite{sakurai2020modern}. The corresponding amplitudes and intensities are
\begin{align}
\tilde\psi_{\pm y}(k_y) &=
\frac{1}{\sqrt{2}}\left(\tilde\psi_\uparrow(k_y) \pm i\,\tilde\psi_\downarrow(k_y)\right), \\[6pt]
I_{\pm y}(y_{\rm scr}) &= \frac{m_e}{\hbar T_{scr}}\,\big|\tilde\psi_{\pm y}(k_y)\big|^2 .
\end{align}

\begin{figure*}
	\centering	\includegraphics[scale=0.9]{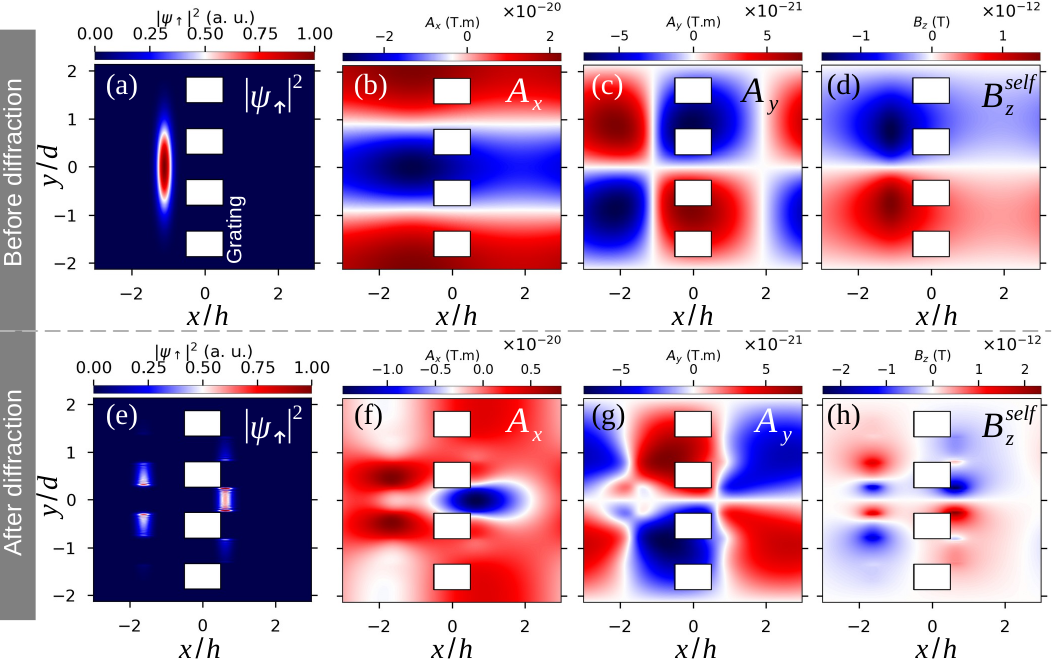}	
	\caption{Simulation results for a spin-up polarized state ($\alpha_0=1$, $\beta_0=0$) in the absence of external magnetic field $\mathbf{B^{ext}}$. Panels (a–d) show the probability density $|\psi_{\uparrow}(x,y,t)|^{2}$ immediately before the grating and the associated self-generated fields: (b) $A_{y}(x,y,t)$, (c) $A_{x}(x,y,t)$, and (d) out-of-plane magnetic field $B^{self}_{z}(x,y,t)$ along $\bf{\hat{z}}$. Positive values of the magnetic field indicate that the field is oriented along the \(\hat{\mathbf{z}}\) direction, whereas negative values indicate orientation along the \(-\hat{\mathbf{z}}\) direction. Panels (e–h) display the corresponding quantities immediately after diffraction: (e) $|\psi_{\uparrow}(x,y,t)|^{2}$, (f) $A_{y}(x,y,t)$, (g) $A_{x}(x,y,t)$, and (h) $B^{self}_z(x,y,t)$.}
    \label{self-field_results}
\end{figure*}
%################################################################
\subsection{Simulation parameters and implementation}

All simulations are carried out within the nonrelativistic Pauli framework, restricted to two dimensions $(x,y)$ with forward propagation along $+\hat{x}$, as described by Eq.~(\ref{eq:PSH}). The initial state is taken as a Gaussian spinor,
\begin{align}
\Psi(x,y,t=0) &= \mathcal N
\exp\!\left[-\frac{(x-x_0)^2}{2\sigma_x^2}
           -\frac{(y-y_0)^2}{2\sigma_y^2}\right] \nonumber \\ & \times
e^{i k_0 (x-x_0)}
\begin{pmatrix}\alpha_0 \\ \beta_0\end{pmatrix},
\label{Eq_initial_wavepacket}
\end{align}
centered at $(x_0,y_0)$ with widths $\sigma_x=5\,\mathrm{nm}$ and $\sigma_y=40\,\mathrm{nm}$. The de Broglie wavelength is $\lambda_{\mathrm{dB}}=2.73\,\text{\AA}$ ($k_0=2\pi/\lambda_{\mathrm{dB}}$), corresponding to an initial kinetic energy $E_0=20$ eV. Here $\mathcal N$ is the normalization constant, defined by $\mathcal{N} = \left[\iint |\Psi(x,y,t=0)|^2 dx\,dy\right]^{-1/2}$, and the coefficients $\alpha_0$ and $\beta_0$ decides the initial spin polarization.

The spatial grid spacing is chosen as \(\Delta x=\Delta y=\lambda_{\mathrm{dB}}/10\) to adequately resolve the de Broglie wavelength. The time step is chosen as \(\Delta t = 9.01\times10^{-18} \,\mathrm{s}\), satisfying the Courant-Friedrichs-Lewy stability condition \cite{5391985}. This choice ensures numerical stability and accurate evaluation of the real-space derivative terms in the split-step scheme while preserving the spectral accuracy of the kinetic propagator. All the simulation parameters are listed in Table~\ref{tab:sim_params}.

\begin{table}[ht]
\caption{\label{tab:sim_params} Simulation parameters used in the calculations.}
\begin{ruledtabular}
\begin{tabular}{lc}
Parameter & Value \\
\hline
de Broglie wavelength, $\lambda_{\mathrm{dB}}$ & $2.73~\text{\AA}$ \\
Electron velocity, $v_x$ & $2.65\times10^6$ m/s \\
Grating periodicity, $d$ & $50~\text{nm}$ \\
Peak amplitude of geometrical potential, $V_0$ & $1200~\text{eV}$ \\
$\sigma_x$ & $5~\text{nm}$ \\
$\sigma_y$ & $40~\text{nm}$ \\
\end{tabular}
\end{ruledtabular}
\end{table}

%#######################################################################
%       Results and Discussions
%#######################################################################
\section{Results and Discussion} \label{result_and_discussions}

\subsection{Self-generated magnetic field}
The self-consistent dynamics of a spin-up polarized state (\(\alpha_0=1,\beta_0=0\)) in the absence of an external magnetic field \(\mathbf{B}^{\mathrm{ext}}\) are shown in Fig.~\ref{self-field_results}. The spin-up probability density \(|\psi_{\uparrow}(x,y,t)|^{2}\), together with the associated self-generated vector potentials \(A^{\mathrm{self}}_{x}\) and \(A^{\mathrm{self}}_{y}\), and the resulting out-of-plane magnetic field \(B^{\mathrm{self}}_{z}\), are shown immediately before (Figs.~\ref{self-field_results}(a)–(d)) and after (Figs.~\ref{self-field_results}(e)–(h)) transmission through the grating. Before the grating (Fig.~\ref{self-field_results}(a)), the wave packet retains its Gaussian profile defined in Eq.~(\ref{Eq_initial_wavepacket}), resulting in smooth and nearly symmetric self-generated field distributions (Fig.~\ref{self-field_results}(d)). The peak magnitude of the out-of-plane self-field is $|B^{\mathrm{self}}_z|\approx 1.5\times10^{-12}$~T, which is several orders of magnitude smaller than the Earth’s magnetic field ($\sim 2.5$–$6.5\times10^{-5}$~T, depending on geographic location) \cite{https://doi.org/10.1029/2019GC008324}.

As the wave packet propagates through the multi-slit grating, diffraction modulates the transmitted wave packet (Fig.~\ref{self-field_results}(e)). This redistribution of the wave packet produces localized current variations near the slit edges, leading to alternating lobes in the self-generated vector potentials \(A_x^{\mathrm{self}}\) and \(A_y^{\mathrm{self}}\) (Figs.~\ref{self-field_results}(f) and (g)). As a consequence, the out-of-plane magnetic field \(B^{\mathrm{self}}_z\) develops dipole- and quadrupole-like structures (Fig.~\ref{self-field_results}(h)), reflecting the diffraction-induced reorganization of the charge current. After transmission through the grating, the peak magnitude of \(B^{\mathrm{self}}_z\) increases to approximately \(\pm 2.4\times10^{-12}\,\mathrm{T}\). These results show that, even in the absence of external magnetic fields, diffraction alone induces structured self-fields through the self-consistent current of the Pauli spinor.

\begin{figure}
	\centering
	\includegraphics[scale=1.0]{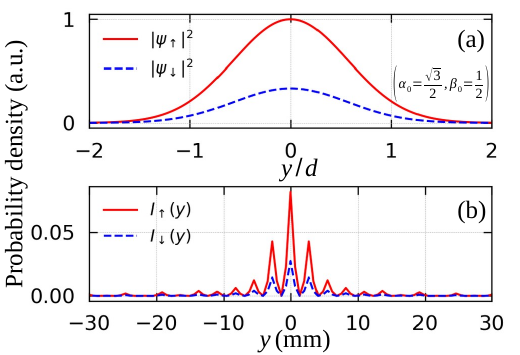}
	\caption{(a) Transverse probability density profiles along the \(y\) direction for the initial spinor components \(|\psi_{\uparrow}(x,y,0)|^{2}\) and \(|\psi_{\downarrow}(x,y,0)|^{2}\), corresponding to the coefficients \(\alpha_{0}=\sqrt{3}/2\) and \(\beta_{0}=1/2\). These values yield initial spin-up and spin-down populations of 75\% and 25\%, respectively. The profiles are normalized to the peak value of the spin-up component. (b) Spin-resolved far-field diffraction intensities \(I_{\uparrow}(y)\) and \(I_{\downarrow}(y)\), recorded at a detection screen placed \(L_{\mathrm{GS}}=50~\mathrm{cm}\) downstream of the nanograting.}
    \label{pattern_pmz_pol_without_B1_B2}
\end{figure}

\begin{figure}
	\centering
	\includegraphics[scale=0.95]{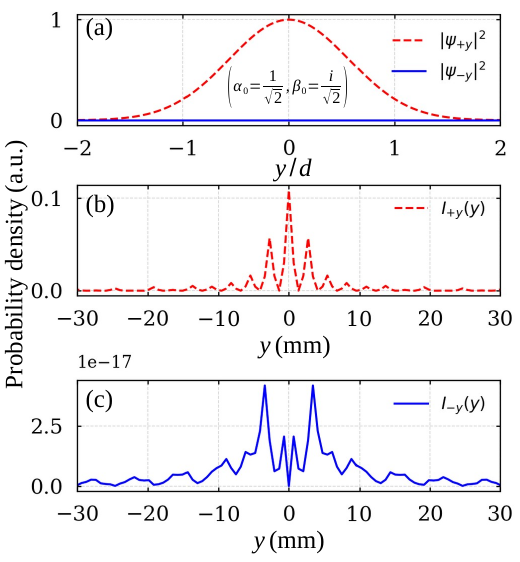}
	\caption{(a) Transverse probability density profiles of the spinor components in the $\sigma_y$ basis, $|\psi_{+y}(x,y,t)|^{2}$ and $|\psi_{-y}(x,y,t)|^{2}$, for an initial state with $\alpha_0 = 1/\sqrt{2}$ and $\beta_0 = i/\sqrt{2}$, corresponding to a pure $+y$-polarized spin state. The profiles are normalized to the peak value of $|\psi_{+y}|^{2}$. Panels (b) and (c) show the corresponding spin-resolved far-field diffraction intensities $I_{+y}(y)$ and $I_{-y}(y)$, respectively, evaluated at a detection screen located $50~\mathrm{cm}$ downstream of the nanograting.}
\label{pattern_pmy_pol_without_B1_B2}
\end{figure}

\begin{figure*}
	\centering
	\includegraphics[scale=0.95]{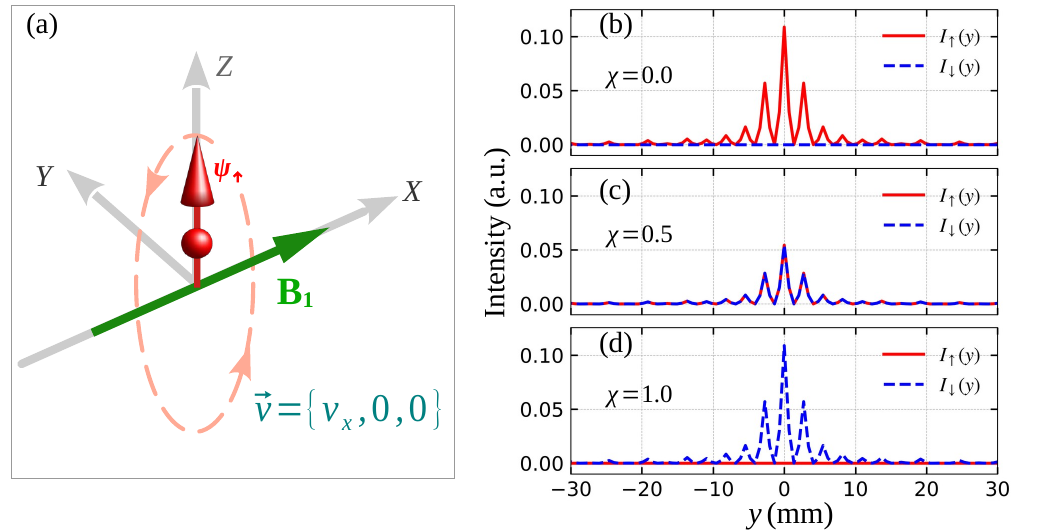}
    \caption{(a) Larmor precession of an initially spin-up polarized electron in a uniform magnetic field $\mathbf{B}_{1}=B_{1}\hat{x}$, with spin rotation about $\hat{x}$ in the $y$–$z$ plane. (b–d) Spin-resolved far-field intensity at a screen located $L_{GS}=50$ cm downstream of the grating for different values of the field ratio $\chi=B_{1}/B_{\pi}$, where $B_{\pi}$ corresponds to a $\pi$ rotation over the interaction region: (b) $\chi=0.00$, (c) $\chi=0.50$, and (d) $\chi=1.00$. The curves show the diffracted spin-up $I_{\uparrow}(y)$ and spin-down $I_{\downarrow}(y)$ components.}
    \label{larmour_precession_and_effect_of_B1}
\end{figure*}

%$$$$$$$$$$$$$$$$$$$$$$
\subsection{Diffraction in the absence of external fields}
To investigate the role of the initial spin state in the far-field diffraction pattern, simulations were performed for different spin populations by varying the coefficients \((\alpha_0,\beta_0)\). Figure~\ref{pattern_pmz_pol_without_B1_B2}(a) shows the transverse probability density profiles of the initial spin-up and spin-down components, \( |\psi_{\uparrow}|^2 \) and \( |\psi_{\downarrow}|^2 \), for \(\alpha_0=\sqrt{3}/2\) and \(\beta_0=1/2\), corresponding to spin populations of 75\% and 25\%, respectively. The corresponding far-field diffraction pattern, obtained using Eq.~(\ref{eq:I_scr_y}), is shown in Fig.~\ref{pattern_pmz_pol_without_B1_B2}(b). The spin-resolved intensities \(I_{\uparrow}(y)\) and \(I_{\downarrow}(y)\) on the detection screen preserve the initial spin populations, demonstrating that the diffraction process does not modify the spin composition. Both spin components exhibit identical fringe widths, \(\gamma = 2.73~\mathrm{mm}\), consistent with the far-field estimate \(\gamma \approx \lambda_{\mathrm{dB}} L_{GS}/d\) \cite{RevModPhys.81.1051}.
The total transmission coefficient, \(T_c=\int_{x=0}^{\infty}\!\int_{y=-\infty}^{\infty} (|\psi_{\uparrow}|^2+|\psi_{\downarrow}|^2)\,dy\,dx\), is found to be \(T_c \approx 0.499\).

Similar behavior is observed for other choices of \((\alpha_0,\beta_0)\), indicating that, in the absence of external magnetic fields, the diffraction pattern directly reflects the initial spin composition and the grating acts as a spin-independent optical element. This can be understood from the two-dimensional geometry of the system. The self-generated magnetic field points along the \(z\)-axis (along \(+\mathbf {\hat{z}}\) for \(y<0\) and along \(-\mathbf {\hat{z}}\) for \(y>0\)). As a result, spins that are aligned with the \(z\)-axis are not affected by this field. In contrast, spinors initialized along \(\hat{\mathbf y}\) are transverse to the self-field direction and are thus maximally sensitive to any self-field–induced spin rotation.

To probe this effect, simulations were performed for an incident spinor initialized in the $\sigma_y$ basis, with $\alpha_0=1/\sqrt{2}$ and $\beta_0=i/\sqrt{2}$. Figure~\ref{pattern_pmy_pol_without_B1_B2}(a) shows the transverse profiles of the probability densities $|\psi_{+y}|^2$ and $|\psi_{-y}|^2$ of both spin components before diffraction, while Figs.~\ref{pattern_pmy_pol_without_B1_B2}(b) and (c) show the corresponding far-field intensities on the screen. Although the initial $-y$ component was absent (Fig.~\ref{pattern_pmy_pol_without_B1_B2}(a)), a finite $I_{-y}(y)$ arises in the far-field pattern (Fig.~\ref{pattern_pmy_pol_without_B1_B2}(c)), which indicates weak spin mixing induced by the self-field. The spin-flip probability from the $+y$ to the $-y$ state is defined as \cite{SHOPE197695},
\begin{align}
P_{+y \rightarrow -y}
&= \frac{\sum_y I_{-y}(y)}
{\sum_y \left[ I_{+y}(y) + I_{-y}(y) \right]}.
\label{eq:P_py_ny}
\end{align}
The computed value of $P_{+y \rightarrow -y}$ is $\sim 1.2 \times 10^{-15}$, which is negligibly small. This confirms that, under free-space propagation without external fields, the self-field is too weak to produce a measurable change in spin dynamics. The grating, therefore, acts as a spin-conserving optical element, preserving the incident spin orientation to high accuracy.

%$$$$$$$$$$$$$$$$$$$$$$$$$$$$$$$$$$
\subsection{Magnetic control of spin-resolved diffraction}
To investigate the effect of an external magnetic field, a uniform field \(\mathbf{B}_{1} = B_1 \hat{\mathbf{x}}\) is applied before the grating, as shown in Fig.~\ref{schematic_diagram_of_simulation_setup} (b), and simulations are performed for different field strengths \(B_1\). In the presence of \(\mathbf{B}_{1}\), the spinor $\Psi(x,y,t)$ undergoes Larmor precession, leading to coherent mixing between the spin components. This process is illustrated in Fig.~\ref{larmour_precession_and_effect_of_B1}(a), where an initially spin-up polarized electron (\(\alpha_0=1\), \(\beta_0=0\)) rotates about the \(\hat{\mathbf{x}}\) axis in the \(y\)–\(z\) plane. For sufficiently strong fields, a complete spin flip can occur. The magnetic field required to induce a \(\pi\) rotation over an interaction length \(L_{B1}\) is given by
\begin{equation}
B_\pi = \frac{4\pi^2 \hbar}{g e L_{B1} \lambda_{\mathrm{dB}}},
\end{equation}
which yields \(B_\pi \simeq 4.76\times10^{-4}\,\mathrm{T}\) for the parameters considered here. For comparison across different field strengths, a dimensionless parameter is introduced as
\begin{equation}
\chi = \frac{B_1}{B_\pi},
\end{equation}
which quantifies the applied magnetic field relative to the characteristic \(\pi\)-rotation field $B_\pi$.

Figures~\ref{larmour_precession_and_effect_of_B1}(b)–(d) show the spin-resolved far-field diffraction intensities \(I_{\uparrow}(y)\) and \(I_{\downarrow}(y)\) for an initially spin-up polarized state (\(\alpha_0=1\), \(\beta_0=0\)) at \(\chi=0\), \(0.5\), and \(1\), respectively. For \(\chi=0\), the diffraction pattern is dominated by the spin-up channel, as shown in Fig.~\ref{larmour_precession_and_effect_of_B1}(b). At \(\chi=0.5\), corresponding to a \(\pi/2\) rotation, the two spin components acquire nearly equal intensities (Fig.~\ref{larmour_precession_and_effect_of_B1}(c)), which indicates spin mixing. For \(\chi=1\), corresponding to a full \(\pi\) rotation, the diffraction population is largely transferred to the spin-down channel, as shown in Fig.~\ref{larmour_precession_and_effect_of_B1}(d).

In all cases, the peak positions and fringe width of the diffraction pattern remain unchanged, while the population is redistributed between the spin channels. This demonstrates that \(\mathbf{B}_{1}\) acts on the spin degree of freedom. Furthermore, as $\chi$ is varied from 0 to 1, the fringe visibility remains unchanged, as evident from the diffraction pattern in Figs.~\ref{larmour_precession_and_effect_of_B1}(b)–(d). The uniform magnetic field $\mathbf{B}_1$ introduces a spatially uniform phase across the transverse ($y$) direction of the wave packet and therefore does not modify the envelope of the diffraction pattern. Therefore, $\mathbf{B}_1$ provides coherent control of population between the spin channels through Larmor precession.

\begin{figure}
    \centering    \includegraphics[scale=0.95]{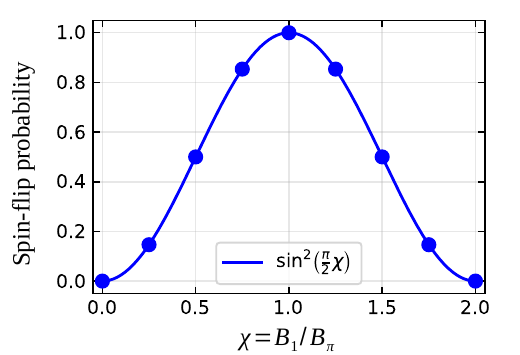}
    \caption{Spin-flip probability $P_{\uparrow\rightarrow\downarrow}$ as a function of the field ratio $\chi = B_{\mathrm{1}}/B_{\pi}$ for an initially spin-up polarized state ($\alpha_0=1$, $\beta_0=0$). Symbols denote simulation results, and the solid line shows the analytic dependence $P_{\uparrow\rightarrow\downarrow}=\sin^{2}\!\left(\tfrac{\pi}{2}\chi\right)$.}
    \label{spin-flip_probability}
\end{figure}

\begin{figure}
	\centering	\includegraphics[scale=0.95]{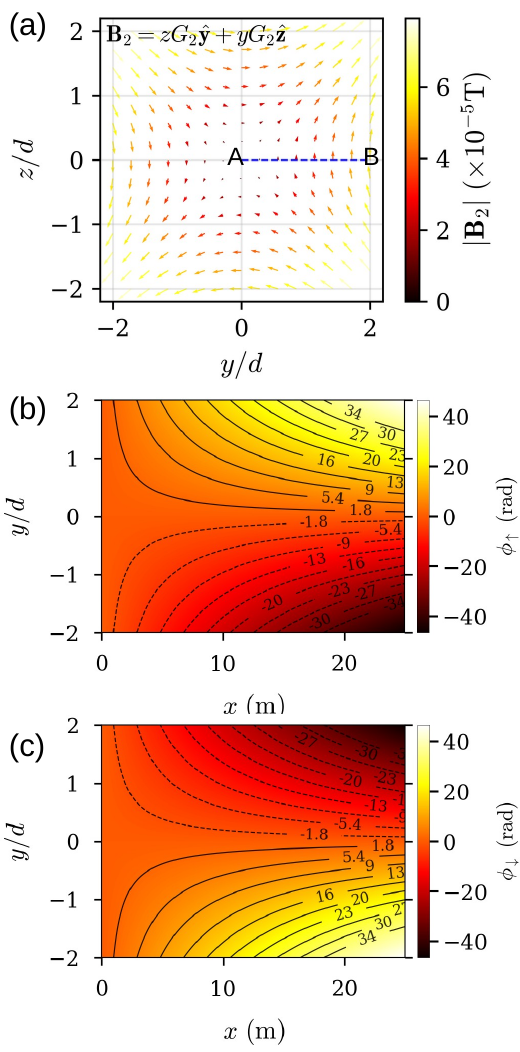}	
	\caption{(a) Profile of the magnetic field $\mathbf{B}_2 = z G_2 \hat{\mathbf y} + y G_2 \hat{\mathbf z}$ in the $yz$ plane. The post-grating magnetic field $\mathbf{B}_2$ induces an accumulated Zeeman phase $\phi_{\uparrow,\downarrow}$ and a spin-dependent transverse deflection $\Delta y_{\mathrm{scr}}$. (b,c) Accumulated Zeeman phase $\phi_{\uparrow}(x,y)$ and $\phi_{\downarrow}(x,y)$ (Eq.~(\ref{zeeman_phase_B2})) acquired over the interaction region $0 \le x \le L_{B2}$. Solid (dashed) contours denote positive (negative) phase. }
    \label{zeeman_phase_and_deflection}
\end{figure}

To quantify the dependence of the spin dynamics on the external magnetic field, the spin-flip probability from spin-up to spin-down, \(P_{\uparrow\rightarrow\downarrow}\), is evaluated using Eq.~(\ref{eq:P_py_ny}). Figure~\ref{spin-flip_probability} shows \(P_{\uparrow\rightarrow\downarrow}\) as a function of \(\chi\) for an initially spin-up polarized electron. The numerical results follow the analytic expression \(P_{\uparrow\rightarrow\downarrow}=\sin^{2}\!\left(\tfrac{\pi}{2}\chi\right)\), which is obtained from unitary spinor rotation due to Larmor precession in a transverse magnetic field \cite{sakurai2020modern}. The value of \(P_{\uparrow\rightarrow\downarrow}\) reaches unity at \(\chi=1\), corresponding to a \(\pi\) rotation, and vanishes at \(\chi=2\), corresponding to a full \(2\pi\) rotation that restores the initial spin state, as shown in Fig.~\ref{spin-flip_probability}. At \(\chi=0.5\), equal spin-up and spin-down populations are obtained. These results demonstrate coherent spin rotation induced by the applied magnetic field \(\mathbf{B}_1\). The spatial separation of the spin components is then investigated by applying a second nonuniform magnetic field \(\mathbf{B}_2\) downstream of the grating.

%#####################################################################
\subsection{Coherent spin filtering by the post-grating magnetic field \texorpdfstring{$\mathbf{B}_2$}{B2}}

To achieve spatial separation of the spin components after diffraction, a nonuniform magnetic field is applied downstream of the grating. The post-grating field is taken in the form
\(\mathbf{B}_2 = z G_2 \hat{\mathbf y} + y G_2 \hat{\mathbf z}\),
with gradient \(G_2 = 560~\mathrm{T/m}\) and an adjustable interaction length \(L_{B2}\), as illustrated in Fig. \ref{schematic_diagram_of_simulation_setup}(b). During interaction with $\mathbf{B}_2$, the transmitted spinor wave function acquires a spatially dependent Zeeman phase, which imparts opposite transverse momentum shifts to the two spin components and enables coherent spin filtering at the detection plane.

In the present two-dimensional simulation, only the dominant transverse magnetic-field component
\(\mathbf{B}_2 \simeq y G_2 \hat{\mathbf z}\) is retained, while the \(z G_2 \hat{\mathbf y}\) component is neglected. This choice confines the electron dynamics to the \(xy\) plane. Inclusion of the \(\hat{\mathbf y}\)-directed magnetic-field component would produce a Lorentz force along the \(z\) direction, leading to out-of-plane motion that cannot be captured within a two-dimensional framework. A two-dimensional profile of \(\mathbf{B}_2 = z G_2 \hat{\mathbf y} + y G_2 \hat{\mathbf z}\) in the perpendicular (\(yz\)) plane is shown in Fig.~\ref{zeeman_phase_and_deflection}(a).
This approximation corresponds to the local magnetic field experienced by the electron beam near the symmetry plane (AB dotted line) and represents the linearized form of a quadrupole-like magnetic field.

For an electron propagating with longitudinal velocity $v_x$, the interaction time within the post-grating magnetic region is $T_{B2}=L_{B2}/v_x$. During this interval, each spin component acquires a position-dependent Zeeman phase due to the interaction with $\mathbf{B_2}$ \cite{sakurai2020modern},
\begin{equation}
\phi_{\uparrow,\downarrow}(y)
= -\frac{1}{\hbar}\int_0^{T_{B2}}\boldsymbol{\mu}\!\cdot\!\mathbf{B}_2\,dt
= \mp\frac{g\mu_B}{2\hbar}\,yG_2\,T_{B2},
\label{zeeman_phase_B2}
\end{equation}
where $\boldsymbol{\mu} = -\tfrac{g\mu_B}{2}\boldsymbol{\sigma}$ is the electron magnetic moment. The accumulated Zeeman phase is controlled by the magnetic-field gradient $G_2$ and the interaction time $T_{B2}$. Figures~\ref{zeeman_phase_and_deflection}(b) and (c) show the spatially accumulated Zeeman phase $\phi_{\uparrow}(x,y)$ and $\phi_{\downarrow}(x,y)$ for the two spin components over a post-grating interaction length $L_{B2}=25$~m. Solid (dashed) contours denote positive (negative) phase values. This phase has equal magnitude and opposite sign for the two spin states and is imparted on the transmitted spinor components $\psi_{\uparrow}(y)$ and $\psi_{\downarrow}(y)$, from which the corresponding far-field intensity distributions $I_{\uparrow,\downarrow}(y_{\mathrm{scr}})$ are obtained.

The spatially varying Zeeman phase $\phi_{\uparrow,\downarrow}(y)$ induces opposite transverse momentum shifts for the two spin components, resulting in a spin-dependent deflection at the detection plane. The magnitude of this deflection at the screen, $\Delta y_{\mathrm{scr}}$, is given by (see Appendix~\ref{appendix_spin-dependent_deflection})
\begin{equation}
\Delta y_{\mathrm{scr}} = \frac{\hbar T_{\mathrm{scr}}}{m_e}
\left(\frac{g\mu_B G_2 L_{B2}}{2\hbar v_x}\right).
\label{B2_diflection_Delta_Y_scr}
\end{equation}
Equation~(\ref{B2_diflection_Delta_Y_scr}) shows that \(\Delta y_{\mathrm{scr}}\) scales linearly with the magnetic-field gradient \(G_2\) and the interaction length \(L_{B2}\). The post-grating magnetic field, therefore, acts as a coherent beam splitter for the spinor components.

\begin{figure} [b]
    \centering    \includegraphics[scale=1.0]{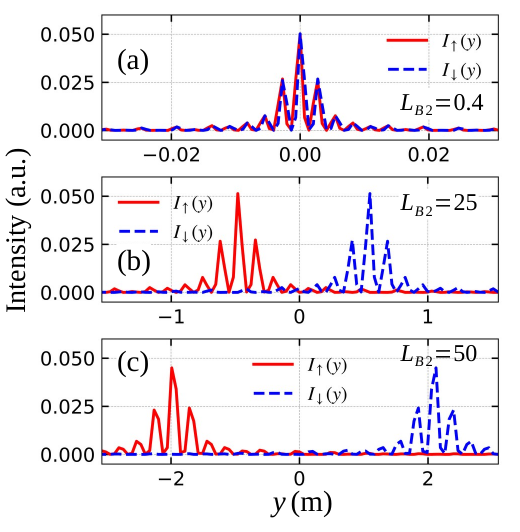}
    \caption{Spin-resolved far-field intensity distributions for an initial state $\alpha_0=\beta_0=1/\sqrt{2}$ in the absence of the upstream field ($\chi=0$). (a--c) Increasing interaction lengths of the post-grating field $\mathbf{B}_2$, $L_{B2}=0.4$, $25$, and $50~\mathrm{m}$, respectively. Larger $L_{B2}$ increases the spin-dependent deflection and yields spatial separation of the spin components.}
    \label{Effect_of_B2_L_B2_Optimization}
\end{figure}

 \begin{figure}
    \centering
    \includegraphics[scale=0.95]{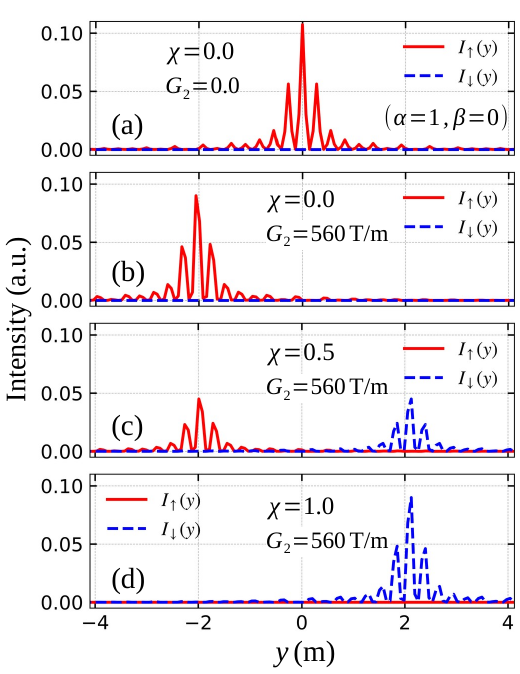}
    \caption{Spin-resolved far-field diffraction patterns for an initially spin-up state ($\alpha=1$, $\beta=0$). (a) Field-free case ($B_1=0$, $G_2=0$). (b--d) Increasing upstream field strengths $\chi=0.0$, $0.5$, and $1.0$ in the presence of a post-grating field gradient $G_2=560~\mathrm{T/m}$. Increasing $\chi$ redistributes the diffraction intensity from the spin-up to the spin-down component while preserving the fringe positions.}

    \label{Effect_of_B2_Spin_Filter}
\end{figure}

We now analyze how \(\mathbf{B}_2\) enables coherent spin filtering by investigating the dependence of the spin-resolved diffraction pattern on the interaction length \(L_{B2}\) and on the combined action of the upstream spin-rotation field \(\mathbf{B}_1\).

\subsubsection{Dependence on interaction length $L_{B2}$}

The effect of the interaction length $L_{B2}$ was investigated by varying it from $0.4~\mathrm{m}$ to $50~\mathrm{m}$ for an initial spinor state $\alpha_0=\beta_0=1/\sqrt{2}$, corresponding to equal populations in the two spin channels, in the absence of the upstream field ($\chi=0$), as shown in Fig.~\ref{Effect_of_B2_L_B2_Optimization}. The simulations were performed with a magnetic-field gradient $G_2=560~\mathrm{T/m}$. For short interaction lengths ($L_{B2}\!<\!1~\mathrm{m}$), the accumulated Zeeman phase is small, and the spin components nearly overlap, as shown in Fig.~\ref{Effect_of_B2_L_B2_Optimization}(a). At intermediate lengths ($L_{B2}\!\approx\!25~\mathrm{m}$), partial separation of the diffraction maxima is observed, as shown in Fig.~\ref{Effect_of_B2_L_B2_Optimization}(b). For $L_{B2}\!=\!50~\mathrm{m}$, the spin-up and spin-down envelopes become completely resolved, as shown in Fig.~\ref{Effect_of_B2_L_B2_Optimization}(c), confirming the coherent spin-filtering effect of the magnetic slab. Across all cases, the overall diffraction pattern remains unchanged, which indicates that $\mathbf{B}_2$ modifies only the spinor phase without affecting the spatial coherence established by the grating.

\begin{figure*}
	\centering
	\includegraphics[scale=1.0]{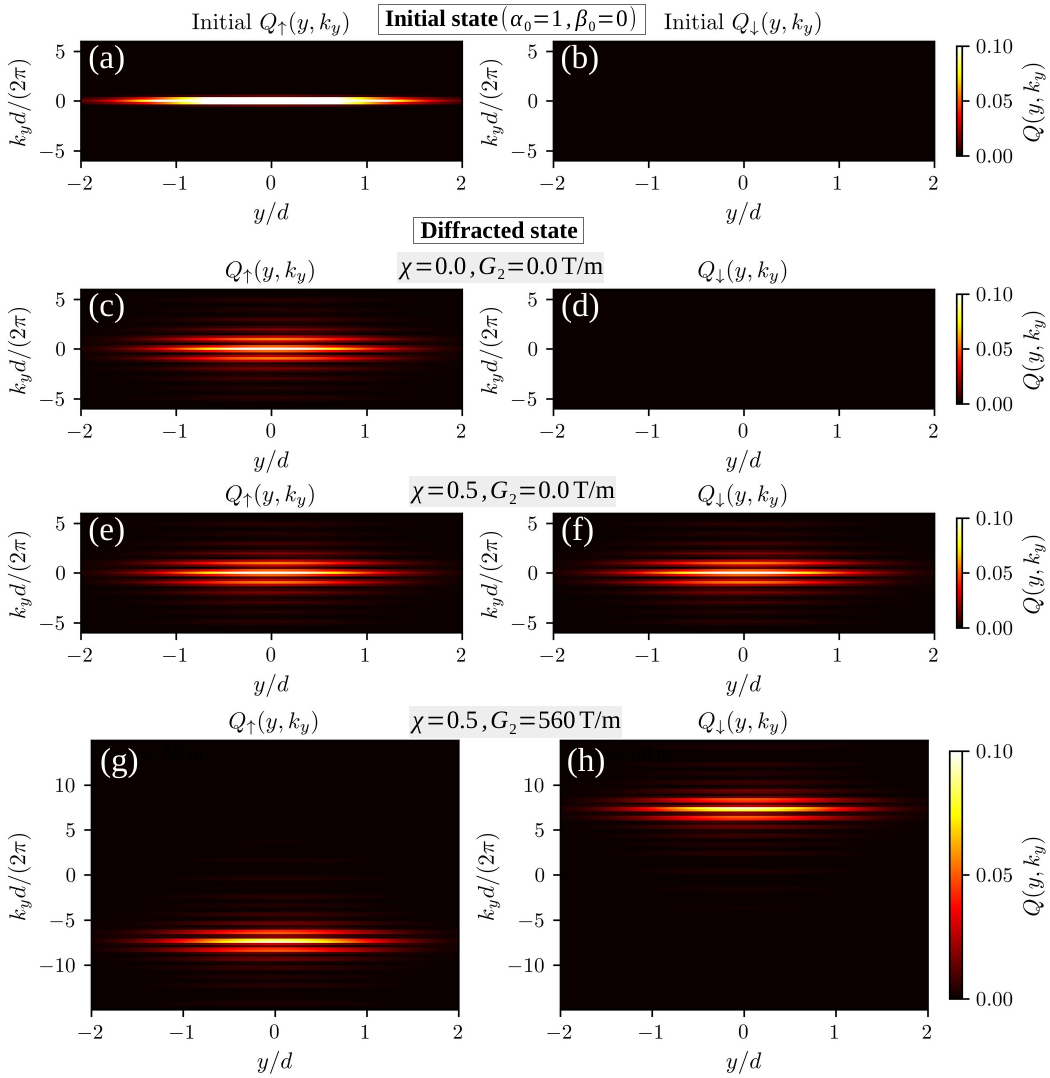}
	\caption{Spin-resolved Husimi $Q(y,k_y)$ maps of the transverse phase space ($y$, $k_y$). (a, b) Initial state before the grating for $(\alpha_0=1, \beta_0=0)$, with the population confined to the spin-up channel. (c, d) Diffracted state in the absence of magnetic fields, showing discrete transverse momentum components. (e, f) State after application of a uniform upstream magnetic field $\mathbf{B_1}$ ($\chi=0.5$), where the population is redistributed between the spin components without a transverse momentum shift. (g, h) State after the downstream magnetic gradient field $\mathbf{B_2}$ ($G_2=560~\mathrm{T/m}$, $L_{B2}=50~\mathrm{m}$), where opposite translations of the Husimi distributions along $k_y$ occur for the two spin components. The preserved distribution shapes indicate coherent spin-dependent momentum control.}
    \label{Husimi_Q_maps}
\end{figure*}

 \begin{figure}
    \centering
    \includegraphics[scale=0.95]{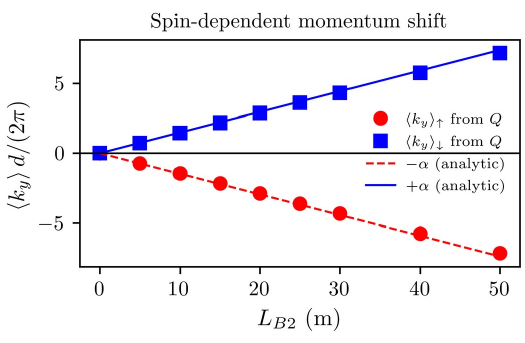}
    \caption{Mean transverse wave number $\langle k_y\rangle$ obtained from the Husimi $Q$-function (Eq. (\ref{eq:kyMean})) for the spin-up and spin-down components as a function of the downstream magnetic-field length $L_{B2}$. Equal and opposite momentum shifts are induced for the two spin components and increase linearly with $L_{B2}$. Symbols represent numerical results, while dashed lines show the analytic prediction $\pm\alpha$ (Eq. (\ref{pm_alpha})). This demonstrates quantitative agreement and coherent spin-dependent momentum control.}    \label{Spin_dependent_momentum_shift}
\end{figure}

\subsubsection{Spin filtering and control of spin populations}

Figure~\ref{Effect_of_B2_Spin_Filter}(a--d) shows the spin-resolved far-field diffraction patterns for an initially spin-up state ($\alpha=1$, $\beta=0$) under different configurations of $\mathbf{B_1}$ and $\mathbf{B_2}$. In the absence of external fields ($B_1=0$, $G_2=0$), the diffraction pattern corresponds to the field-free case and remains entirely in the spin-up channel, as shown in Fig.~\ref{Effect_of_B2_Spin_Filter}(a).

When the nonuniform post-grating field $\mathbf{B}_2$ is applied together with the upstream uniform field $\mathbf{B}_1$, the diffraction pattern becomes spin dependent. The field $\mathbf{B}_1$ induces coherent spin rotation through Larmor precession, while $\mathbf{B}_2$ imprints a spatially dependent Zeeman phase that separates the rotated spin components at the detection plane, as seen in Figs.~\ref{Effect_of_B2_Spin_Filter}(b--d).

As the strength of $\mathbf{B}_1$ is increased, the population is continuously redistributed between the spin-up and spin-down diffraction channels without altering the fringe positions or width $\gamma$. The spatial separation of the two spin components remains controlled by $\mathbf{B}_2$.

These results demonstrate $\mathbf{B}_1$ as a coherent spin 
population controller and $\mathbf{B}_2$ as a spin-dependent spatial analyzer. Their combined action enables tunable spin filtering in free-electron diffraction while preserving spatial coherence.

%#####################################################################
\subsection{Spin-resolved Husimi phase-space maps}
\label{sec:Husimi}

To obtain a direct phase-space representation of the spin-dependent transverse dynamics, we employ the Husimi $Q$-function constructed from the transmitted spinor wave function. The Husimi representation provides a smooth, positive-definite phase-space distribution in the transverse coordinate $y$ and its conjugate wave number $k_y$. This enables visualization of spin-dependent momentum shifts induced by magnetic fields.

For each spin component $s$, the Husimi $Q$-function is defined as \cite{LEE1995147}
\begin{equation}
Q_s(y_0,k_{y0}) = \frac{1}{\pi}
\left|
\int dy\,
\phi^{*}_{y_0,k_{y0}}(y)\,
\psi_s(y)
\right|^2 ,
\label{eq:HusimiQ}
\end{equation}
where $\psi_s(y)$ is the transverse wave function evaluated at a specified axial position $x$ (taken either at $x=0$ or at the detection screen), and $\phi_{y_0,k_{y0}}(y)$ is a minimum-uncertainty Gaussian probe state,
\begin{equation}
\phi_{y_0,k_{y0}}(y) =
\left(\frac{1}{\pi\sigma^2}\right)^{1/4}
\exp\!\left[-\frac{(y-y_0)^2}{2\sigma^2}\right]
\exp\!\left[i k_{y0}(y-y_0)\right].
\label{eq:HusimiProbe}
\end{equation}
The width $\sigma$ is chosen equal to the transverse width of the incident wave packet, which provides a physically motivated and fixed phase-space resolution that satisfies the uncertainty relation $\Delta y\,\Delta k_y = 1/2$. The qualitative features discussed below are insensitive to moderate variations of $\sigma$ around this value.

Figure~\ref{Husimi_Q_maps} shows the spin-resolved Husimi phase-space distributions for different cases of magnetic manipulation. Before diffraction, the phase-space distribution is localized near $k_y=0$ and confined to the spin-up channel, corresponding to the initial Gaussian wave packet (Figs.~\ref{Husimi_Q_maps}(a) and (b)). After transmission through the grating in the absence of magnetic fields, discrete transverse momentum components emerge due to diffraction, while the Husimi distribution for the spin-up channel remains centered at $k_y=0$, as shown in Figs.~\ref{Husimi_Q_maps}(c) and (d). The preservation of the phase-space structure of the initial state (Figs.~\ref{Husimi_Q_maps}(a) and (b)) for both spin components confirms spin-independent coherent diffraction in the absence of magnetic fields.

The application of a uniform upstream magnetic field \(\mathbf{B}_{1}\) redistributes the population between the spin components through coherent Larmor rotation without modifying the transverse phase-space structure or inducing a transverse momentum shift, as shown in Figs.~\ref{Husimi_Q_maps}(e) and (f). In contrast, the downstream magnetic gradient field \(\mathbf{B}_{2}\) produces opposite translations of the Husimi distributions along the $k_y$ axis for the two spin components (Figs.~\ref{Husimi_Q_maps}(g) and (h)). The absence of additional broadening or distortion during this translation indicates that \(\mathbf{B}_{2}\) induces coherent, spin-dependent transverse momentum shifts while preserving the diffraction-induced phase-space structure.

A quantitative measure of the spin-dependent momentum shift is obtained from the first moment of the Husimi distribution \cite{LEE1995147},
\begin{equation}
\langle k_y \rangle_s =
\frac{\iint k_y\, Q_s(y,k_y)\, dy\, dk_y}
     {\iint Q_s(y,k_y)\, dy\, dk_y},
\label{eq:kyMean}
\end{equation}
which provides a centroid-based diagnostic that is insensitive to fine-scale diffraction oscillations and to the overall normalization of $Q_s$.

Figure~\ref{Spin_dependent_momentum_shift} shows the dependence of the mean transverse wave number $\langle k_y\rangle$ on the interaction length $L_{B2}$ of the downstream magnetic gradient field. The two spin components acquire equal-magnitude but opposite momentum shifts that increase linearly with $L_{B2}$. The numerical results obtained from Eq.~(\ref{eq:kyMean}) follow the analytic prediction $\langle k_y\rangle_{\uparrow,\downarrow}=\pm\alpha$ (Eq.~(\ref{pm_alpha})) over the full range of $L_{B2}$. This shows quantitative agreement between simulation and theory. This linear scaling is consistent with the Zeeman phase gradient imprinted by $\mathbf{B}_{2}$ and allows the spin-dependent momentum shift to be characterized in phase space using the Husimi $Q$-function.

%\textbf{\textcolor{blue}{Discussions:}}
%===========================================================================
%                Conclusion
%===========================================================================
\section{Conclusion}\label{conclusions}

A self-consistent Maxwell-Pauli model of spin-resolved electron diffraction from nanogratings has been developed. Image-charge interactions, self-generated magnetic fields, and externally applied magnetic fields are included. Quantitative simulations show that the magnetostatic self-field associated with the electron probability current is several orders of magnitude too weak to induce measurable spin mixing. Nanogratings therefore act as spin-conserving beam splitters under field-free conditions.

Coherent spin control is achieved by applying an external magnetic field $\mathbf{B}_1$ before the diffraction grating, which induces Larmor precession of the electron spin. The characteristic $\pi$-rotation field $B_\pi$ scales inversely with the interaction length and the electron de Broglie wavelength. This scaling demonstrates the high magnetic sensitivity of low-energy electrons, which can be used as a magnetic field sensor. For example, if we consider $\mathbf{B_1}$ to change from $\mathbf{B_1}=B_{\pi} \simeq 4.76 \times 10^{-4}\,\mathrm{T}$ to $\mathbf{B_1}\pm\Delta \mathbf{B_1}$ (where $\Delta \mathbf{B_1}=0.01 \mathbf{B_1}$), then $\chi$ changes by $1\%$, resulting in a change in the spin population by $1\%$. The sensitivity may be further improved by reducing the value of $B_{\pi}$ through $L_{B1}$. A downstream nonuniform magnetic field $\mathbf{B}_2$ imparts a spatially varying Zeeman phase on the diffracted wave. Opposite transverse momentum shifts are induced for the two spin components. Tunable spatial separation of the spin-resolved diffraction envelopes is thereby obtained. Fringe periodicity and spatial coherence are preserved.

The spin-dependent transverse dynamics are further quantified using spin-resolved Husimi $Q$-function phase-space maps. This phase-space analysis shows that $\mathbf{B}_1$ redistributes population between spin components without altering the transverse momentum structure, whereas $\mathbf{B}_2$ produces equal-magnitude and opposite translations of the Husimi distributions along the transverse wave-number axis. The invariant phase-space structure provides a direct confirmation that the induced spin-dependent momentum shifts are coherent and consistent with analytic predictions.

The combined action of $\mathbf{B}_1$ and $\mathbf{B}_2$ enables precise control over the spin populations (up and down) within the separated components, which enables spin-resolved free-electron generation and interferometry. It also provides a quantitative basis for spin-contrast electron diffraction and magnetic-field sensing through measurable spin-dependent fringe shifts at low electron energies.

%&&&&&&&&&&&&&&&&&&&&&&&&&&&&&&&&&&&&&&&&&&&&&&&&&&&&&&&&&&&&&&&&&&&&&&&&&&&&&&&&&&&&&&&&&&&&&&&&&&&&&&&&&&&&&&&&&&&&&&&&&&&&&&&&&&&&&&&&&&&&&&&&&&&&&&&&&&&&&&&&&&&&&&&&&&&&&&&
%#####################################################################
\appendix

\section{Explicit expansion of the Hamiltonian in the uniform magnetic field $\mathbf{B}_1$ employed before the grating}
\label{app:PauliExpansion_B1}

The Hamiltonian $\mathbf{H}_1$ governing the spinor evolution in the presence of the uniform magnetic field
$\mathbf{B}_1 = B_1\,\hat{\mathbf x}$, the associated vector potential
$\mathbf{A}_{1}= y B_1\,\hat{\mathbf z}$, and the self-generated vector potential
$\mathbf{A}^{\mathrm{self}}$ is given by
\begin{equation}
\mathbf{H}_1
=
\frac{1}{2m_e}
\left(
\mathbf p
+
e\,\mathbf A_1
+
e\,\mathbf A^{\mathrm{self}}
\right)^2
-
\frac{g\mu_B}{2}\,
\boldsymbol{\sigma}\cdot\mathbf B_1 ,
\label{eq:H1_app_start}
\end{equation}
where $\mathbf p=-i\hbar\nabla$ is the canonical momentum operator and
$\boldsymbol{\sigma}$ denotes the Pauli matrices.

Since the vector potential $\mathbf A_1$ has only a $z$ component and the dynamics
is restricted to the $xy$ plane, $\mathbf A_1$ does not contribute to the transverse
kinetic terms. Accordingly, the Hamiltonian reduces to
\begin{equation}
\mathbf{H}_1
=
\frac{1}{2m_e}
\left(
\mathbf p
+
e\,\mathbf A^{\mathrm{self}}
\right)^2
-
\frac{g\mu_B}{2}\,
\boldsymbol{\sigma}\cdot\mathbf B_1 .
\label{eq:H1_app_compact}
\end{equation}

Expanding the kinetic energy term (first term) in Eq.~(\ref{eq:H1_app_compact}),
\begin{equation}
\left(
\mathbf p
+
e\,\mathbf A^{\mathrm{self}}
\right)^2
=
\mathbf p^2
+
e\left(
\mathbf p\cdot\mathbf A^{\mathrm{self}}
+
\mathbf A^{\mathrm{self}}\cdot\mathbf p
\right)
+
e^2
\left(\mathbf A^{\mathrm{self}}\right)^2 .
\end{equation}
Substituting this expression into Eq.~(\ref{eq:H1_app_compact}) yields
\begin{align}
\mathbf{H}_1
&=
\frac{\mathbf p^2}{2m_e}
+
\frac{e}{2m_e}
\left(
\mathbf p\cdot\mathbf A^{\mathrm{self}}
+
\mathbf A^{\mathrm{self}}\cdot\mathbf p
\right) 
+
\frac{e^2}{2m_e}
\left(\mathbf A^{\mathrm{self}}\right)^2
\nonumber \\ &
-
\frac{g\mu_B}{2}\,
\boldsymbol{\sigma}\cdot\mathbf B_1 .
\label{eq:H1_app_expand}
\end{align}

The first term in Eq. (\ref{eq:H1_app_expand}) corresponds to the free-particle kinetic energy,
\begin{equation}
\frac{\mathbf p^2}{2m_e}
=
-\frac{\hbar^2}{2m_e}\nabla^2 .
\end{equation}

The Zeeman interaction term (last term) in Eq. (\ref{eq:H1_app_expand}) with the uniform magnetic field $\mathbf B_1 = B_1 \hat{\mathbf x}$
takes the explicit form
\begin{equation}
-\frac{g\mu_B}{2}\,
\boldsymbol{\sigma}\cdot\mathbf B_1
=
-\frac{g\mu_B}{2}\,B_1\,\sigma_x .
\end{equation}

Collecting all contributions, the Hamiltonian $\mathbf{H}_1$ can be written explicitly as
\begin{align}
\mathbf{H}_1
&=
-\frac{\hbar^2}{2m_e}\nabla^2
+
\frac{e}{2m_e}
\left(
\mathbf p\cdot\mathbf A^{\mathrm{self}}
+
\mathbf A^{\mathrm{self}}\cdot\mathbf p
\right) \nonumber\\ &
+
\frac{e^2}{2m_e}
\left(\mathbf A^{\mathrm{self}}\right)^2
-
\frac{g\mu_B}{2}\,B_1\,\sigma_x .
\label{eq:H1_app_final}
\end{align}

%#################################################################################
\section{Explicit expansion of the Hamiltonian in the nonuniform magnetic field $\mathbf{B}_2$ employed after the grating}
\label{app:PauliExpansion_B2}

The Hamiltonian $\mathbf{H}_2$ governing the spinor evolution in the presence of the nonuniform magnetic field
$\mathbf{B}_2 = yG_2\,\hat{\mathbf z}$, the associated vector potential $\mathbf{A}_{2}=-(y^{2}G_{2}/2)\,\hat{\mathbf x}$,
and the self-generated vector potential $\mathbf{A}^{\mathrm{self}}$ is given by
\begin{equation}
\mathbf{H}_2
=
\frac{1}{2m_e}
\left(
\mathbf p
+
e\,\mathbf A_2
+
e\,\mathbf A^{\mathrm{self}}
\right)^2
-
\frac{g\mu_B}{2}\,
\boldsymbol{\sigma}\cdot\mathbf B_2 ,
\label{eq:H2_app_start}
\end{equation}

Defining the total vector potential,
\begin{equation}
\mathbf A_{\mathrm{tot}}=\mathbf A_2+\mathbf A^{\mathrm{self}},
\end{equation}
the Hamiltonian may be written as
\begin{equation}
\mathbf{H}_2
=
\frac{1}{2m_e}
\left(
\mathbf p + e\,\mathbf A_{\mathrm{tot}}
\right)^2
-
\frac{g\mu_B}{2}\,
\boldsymbol{\sigma}\cdot\mathbf B_2 .
\label{eq:H2_compact}
\end{equation}

Expanding the kinetic term (the first term in Eq.~(\ref{eq:H2_compact})),
\begin{equation}
\left(
\mathbf p + e\,\mathbf A_{\mathrm{tot}}
\right)^2
=
\mathbf p^2
+
e\left(
\mathbf p\cdot\mathbf A_{\mathrm{tot}}
+
\mathbf A_{\mathrm{tot}}\cdot\mathbf p
\right)
+
e^2\,\mathbf A_{\mathrm{tot}}^2 .
\end{equation}
Substituting into Eq.~(\ref{eq:H2_compact}) gives
\begin{align}
\mathbf{H}_2
&=
\frac{\mathbf p^2}{2m_e}
+
\frac{e}{2m_e}
\left(
\mathbf p\cdot\mathbf A_{\mathrm{tot}}
+
\mathbf A_{\mathrm{tot}}\cdot\mathbf p
\right)
+
\frac{e^2}{2m_e}
\mathbf A_{\mathrm{tot}}^2 \nonumber \\ &
-
\frac{g\mu_B}{2}\,
\boldsymbol{\sigma}\cdot\mathbf B_2 .
\label{eq:H2_app_expand}
\end{align}

Separating the external and self-field contributions,
\begin{align}
\mathbf p\cdot\mathbf A_{\mathrm{tot}}+\mathbf A_{\mathrm{tot}}\cdot\mathbf p
&=
\mathbf p\cdot\mathbf A_2
+
\mathbf p\cdot\mathbf A^{\mathrm{self}}
+
\mathbf A_2\cdot\mathbf p
+
\mathbf A^{\mathrm{self}}\cdot\mathbf p, \\
\mathbf A_{\mathrm{tot}}^2
&=
\mathbf A_2^2
+
\left(\mathbf A^{\mathrm{self}}\right)^2
+
2\,\mathbf A_2\cdot\mathbf A^{\mathrm{self}} .
\end{align}
so that
\begin{align}
\mathbf{H}_2
&=
\frac{\mathbf p^2}{2m_e}
+
\frac{e}{2m_e}
\left(\mathbf p\cdot\mathbf A_2+\mathbf A_2\cdot\mathbf p\right) \nonumber \\ &
+
\frac{e}{2m_e}
\left(\mathbf p\cdot\mathbf A^{\mathrm{self}}+\mathbf A^{\mathrm{self}}\cdot\mathbf p\right)
+
\frac{e^2}{2m_e} \mathbf A_2^2
\nonumber\\
&
+
\frac{e^2}{2m_e} \left(\mathbf A^{\mathrm{self}}\right)^2
+
\frac{e^2}{m_e} \,\mathbf A_2\cdot\mathbf A^{\mathrm{self}}
-
\frac{g\mu_B}{2}\,
\boldsymbol{\sigma}\cdot\mathbf B_2 .
\label{eq:H2_app_expand_2}
\end{align}

The individual terms in Eq.~(\ref{eq:H2_app_expand_2}) are evaluated below. The first term is
\begin{equation}
\frac{\mathbf p^2}{2m_e}
=
-\frac{\hbar^2}{2m_e}\nabla^2 .
\end{equation}

Since $A_{2x}$ depends only on $y$, one has $[p_x,A_{2x}]=0$, and therefore
\begin{equation}
\mathbf p\cdot\mathbf A_2+\mathbf A_2\cdot\mathbf p
=
2A_{2x}p_x
=
-\;G_2 y^2\,p_x .
\end{equation}
Hence, the second term on the right-hand side of Eq.~(\ref{eq:H2_app_expand_2}) becomes
\begin{equation}
\frac{e}{2m_e}
\left(
\mathbf p\cdot\mathbf A_2+\mathbf A_2\cdot\mathbf p
\right)
=
-\frac{e G_2}{2m_e}\,y^2 p_x
=
\frac{i e\hbar G_2}{2m_e}\,y^2 \partial_x .
\end{equation}

The fourth term on the right-hand side of Eq.~(\ref{eq:H2_app_expand_2}) is
\begin{equation}
\frac{e^2}{2m_e}\mathbf A_2^2
=
\frac{e^2 G_2^2}{8m_e}\,y^4 .
\end{equation}

The sixth term on the right-hand side of Eq.~(\ref{eq:H2_app_expand_2}) is
\begin{equation}
\frac{e^2}{m_e}\mathbf A_2\cdot\mathbf A^{\mathrm{self}}
=
-\frac{e^2 G_2}{2m_e}\,y^2 A_x^{\mathrm{self}} .
\end{equation}

The Zeeman term becomes
\begin{equation}
-\frac{g\mu_B}{2}\boldsymbol{\sigma}\cdot\mathbf B_2
=
-\frac{g\mu_B}{2}\,y G_2\,\sigma_z .
\end{equation}

Collecting all contributions, $\mathbf{H}_2$ takes the explicit form
\begin{align}
\mathbf{H}_2
&=
-\frac{\hbar^2}{2m_e}\nabla^2
-\frac{e G_2}{2m_e}\,y^2 p_x
+\frac{e}{2m_e}
\left(
\mathbf p\cdot\mathbf A^{\mathrm{self}}
+
\mathbf A^{\mathrm{self}}\cdot\mathbf p
\right)
\nonumber\\
&\quad
+\frac{e^2 G_2^2}{8m_e}\,y^4
+\frac{e^2}{2m_e}
\left(\mathbf A^{\mathrm{self}}\right)^2
-\frac{e^2 G_2}{2m_e}\,y^2 A_x^{\mathrm{self}}
-\frac{g\mu_B}{2}\,y G_2\,\sigma_z.
\label{eq:H2_app_final}
\end{align}

\section{Analytic estimate of spin-dependent deflection}\label{appendix_spin-dependent_deflection}

The spin-dependent displacement at the detection screen can be derived from the Zeeman interaction experienced by the electron spin in the post-grating magnetic field $\mathbf{B}_2$. 
For a linear magnetic field $|\mathbf{B_2}|=G_2 y$, the Zeeman phase acquired by the spinor components during the interaction time $T_{B2}=L_{B2}/v_x$ is (Eq. \ref{zeeman_phase_B2})
\begin{equation}
\phi_{\uparrow,\downarrow}(y)
=\mp\frac{g\mu_B}{2\hbar}\,y\,G_2\,T_{B2}
\equiv \mp\alpha \, y,
\qquad
\alpha=\frac{g\mu_B}{2\hbar}\,G_2\,T_{B2},
\label{pm_alpha}
\end{equation}
where $g$ is the electron $g$-factor and $\mu_B$ is the Bohr magneton. The transmitted spinor wave functions after passing through the magnetic region are therefore \cite{PhysRevResearch.6.023165}
\begin{equation}
\psi_{\uparrow,\downarrow}^{\,(\mathrm{after})}(y)
=e^{\mp i\alpha y}\,\psi_{\uparrow,\downarrow}^{\,(\mathrm{0})}(y).
\end{equation}
Using the Fourier shift theorem, this spatially varying phase produces a shift in the transverse momentum domain \cite{bracewell1989fourier},
\begin{equation}
\tilde{\psi}_{\uparrow,\downarrow}^{\,(\mathrm{after})}(k_y)
=\tilde{\psi}_{\uparrow,\downarrow}^{\,(\mathrm{0})}(k_y\pm\alpha),
\end{equation}
which indicates a change in the transverse wave number
\(\Delta k_y = \pm \alpha.\)
Under paraxial propagation to the screen, the transverse position is related to $k_y$ by 
$y_{\mathrm{scr}}=(\hbar k_y/m_e)\,T_{\mathrm{scr}}$, where $T_{\mathrm{scr}}=L_{\mathrm{GS}}/v_x$ is the drift time to the detection plane.
Hence, the spin-dependent shift in position becomes
\begin{equation}
\Delta y_{\uparrow,\downarrow}
=\frac{\hbar T_{\mathrm{scr}}}{m_e}\,\Delta k_y
=\pm\frac{\hbar T_{\mathrm{scr}}}{m_e}\left(\frac{g\mu_B G_2 T_{B2}}{2\hbar}\right).
\end{equation}
Therefore, the magnitude of the spin-dependent deflection is
\begin{equation}
\Delta y
=\frac{\hbar T_{\mathrm{scr}}}{m_e}
\left(\frac{g\mu_B G_2 T_{B2}}{2\hbar}\right),
\end{equation}
which increases linearly with both the magnetic-field gradient $B_2$ and the interaction length $L_{B2}$, since $T_{B2}=L_{B2}/v_x$.

%#######################################

%==========================================
%==========================================
\nocite{*}
\bibliography{bibfile}

\end{document}